\documentclass[12pt]{amsart}
\date{January 13, 2016}

\usepackage{amsmath,amsthm,amsfonts,amssymb,multicol}
\usepackage{graphicx}
\usepackage{booktabs}
\usepackage{array}
\usepackage{subfigure}
\usepackage{enumerate}
\usepackage{url}

\usepackage[english]{babel}
\usepackage{verbatim} 
\usepackage{enumitem}
\usepackage{bbm} 
\usepackage[normalem]{ulem} 
\usepackage{bm}
\usepackage{mathtools}
\usepackage[pdftex]{hyperref}
\usepackage{graphicx}

\newcommand{\be}{\begin}
\newcommand{\e}{\end}
\newcommand{\beq}{\begin{equation}}
\newcommand{\eeq}{\end{equation}}
\newcommand{\beqs}{\begin{equation*}}
\newcommand{\eeqs}{\end{equation*}}
\renewcommand{\l}{\left}
\renewcommand{\r}{\right}

\renewcommand{\d}{\mathrm{d}} 

\newcommand{\set}[1]{\mathbb{#1}}

\newcommand{\curly}[1]{\mathcal{#1}}

\newcommand{\om}{\omega}

\newcommand{\eps}{\epsilon}

\newcommand{\lam}{\lambda}

\newcommand{\de}{\delta}

\newcommand{\A}{\mathcal{A}}

\newcommand{\ind}{\mathbbm{1}}		

\newcommand{\ttmatrix}[4]{\left(\be{array}{cc} #1&#2\\	#3&#4 \e{array}	\right)}

\newcommand{\scp}[2]{\l\langle#1,#2\r\rangle}

\renewcommand{\it}{\infty}

\newcommand{\tr}[1]{\textnormal{tr}\left[#1\right]}	

\newtheorem{thm}{Theorem}[section]

\newtheorem{prop}[thm]{Proposition}

\theoremstyle{definition}

\theoremstyle{remark}

\def\dotuline{\bgroup
  \ifdim\ULdepth=\maxdimen  
   \settodepth\ULdepth{(j}\advance\ULdepth.4pt\fi
  \markoverwith{\begingroup
  \advance\ULdepth0.08ex
  \lower\ULdepth\hbox{\kern.15em .\kern.1em}%
  \endgroup}\ULon}

\def\dashuline{\bgroup
  \ifdim\ULdepth=\maxdimen  
   \settodepth\ULdepth{(j}\advance\ULdepth.4pt\fi
  \markoverwith{\kern.15em
  \vtop{\kern\ULdepth \hrule width .3em}%
  \kern.15em}\ULon}
\allowdisplaybreaks

\let\<\langle
\let\>\rangle

\newcommand{\E}{\mathbb E}
\newcommand{\R}{\mathbb R}
\newcommand{\N}{\mathbb N}

\begin{document}

\title[XY chain with decaying randomness]{On polynomial Lieb-Robinson bounds for the XY chain in a decaying random field} 

\author[M.\ Gebert]{Martin Gebert}
\address{Mathematisches Institut,
  Ludwig-Maximilians-Universit\"at M\"unchen,
  Theresienstra\ss{e} 39,
  80333 M\"unchen, Germany}
\email{gebert@math.lmu.de}

\author[M. Lemm]{Marius Lemm}
\address{Mathematics Dept. MC 253-37, California Institute of Technology, Pasadena, CA 91125}
\email{mlemm@caltech.edu}

\be{abstract}
We consider the isotropic XY quantum spin chain in a random external field in the $z$ direction, with single site distributions given by i.i.d.\ random variables times the critical decaying envelope $j^{-1/2}$. Our motivation is the study of many-body localization. We investigate transport properties in terms of polynomial Lieb-Robinson (PLR) bounds. We prove a zero-velocity PLR bound for large disorder strength $\lam$ and for small $\lam$ we show a partial converse, which suggests the existence of non-trivial transport in the model. 


\e{abstract}

\maketitle

\section{Introduction}
It is well known that a single quantum particle in one dimension which is subjected to an arbitrarily weak random potential exhibits exponential Anderson localization \cite{Anderson, KS}. In the presence of interactions, one enters the subject of many-body localization (MBL) which has been a hot topic of condensed-matter physics in recent years, see e.g.\ \cite{MBL0,MBL5,MBL2,MBL4,MBL1,MBL3} and references therein. On a heuristic level, MBL is described as \emph{absence of thermalization}. Proposed criteria for this include the validity of an area law for the entanglement entropy and absence of information propagation (e.g.\ a zero-velocity Lieb-Robinson bound and logarithmic in time growth of the entanglement entropy). For an extensive list of possible criteria, see the review \cite{MBreview}. The very special MBL phase is expected to break down for sufficiently strong interactions, in what is called the \emph{MBL transition} \cite{MBLT1, MBLT2}. 

A possible starting point for understanding MBL is the XY chain in an i.i.d.\ random field. This is an integrable toy model which can be mapped to non-interacting fermions in a random environment. Since the fermions are then localized in the usual Anderson sense, it can be shown rigorously that this model enjoys an area law for the entanglement entropy for large classes of states \cite{ANSS,AS,PS} and a zero-velocity Lieb-Robinson bound \cite{PhysRevLett.99.167201, MR2966945}. However, a shortcoming of this toy model (apart from integrability) is that it will never display a transition to a non-MBL phase because the fermions are localized at arbitrarily small disorder strength (which is equivalent to arbitrarily large interaction strength).\\

In this paper, we propose a variation of the above toy model which \emph{rigorously displays features suggesting that such a transition occurs} as the disorder strength is varied. The model is the isotropic XY quantum spin chain on the half line with a random and \emph{decaying} external field in the $z$ direction. The Hamiltonian reads
\beqs
    H_{n}^{XY}(\omega):= - \sum_{j=1}^{n-1} \l(\sigma^x_j \sigma^x_{j+1} +\sigma^y_j \sigma^y_{j+1}   \r) +\lambda\sum_{j=1}^n \frac {V_j(\omega)}{j^{1/2}}  \sigma^z_j
\eeqs
where $V_j$ are i.i.d.\ and centered random variables and $\lam>0$ is a parameter describing the disorder strength. Note the decaying envelope $j^{-1/2}$ for the random field. It is critical in that the potential is just barely not in $\ell^2(\N)$. For other decay rates, the random field is either too weak or too strong to observe a qualitative transition from MBL to non-MBL features (such as transport) when $\lam$ is varied.\\

We now explain in which sense our system exhibits features suggesting a transition from transport to localization as the disorder strength $\lam>0$ is \emph{in}creased. While our results will be more general and include bounds on the particle number transport as well, the key notion for quantifying many-body transport for this model are new \emph{anomalous polynomial Lieb-Robinson (PLR) type bounds}. The traditional Lieb-Robinson (LR) bounds, see \cite{LR,NS}, apply to general local Hamiltonians defined on a lattice and establish the existence of a certain ``light cone'' in spacetime outside of which correlations are exponentially small. 

 We say PLR$(a,b)$ holds for parameters $0\leq a\leq 1$ and $b>0$, if there exists a universal constant $C>0$ such that for any observables $A$ supported at site $1$ and $B$ supported at site $k>1$, we have the bound
\beq
\label{eq:polyLR}
\|[\tau_t^n(A),B]\| \leq C \|A\| \|B\|\l(\frac{t^a}{k}\r)^b.
\eeq
Here $\tau_t^N$ is the Heisenberg time evolution generated by the Hamiltonian $H_n^{XY}$, see \eqref{HBdefn}, and $\|\cdot\|$ is the standard operator norm. Intuitively, PLR$(a,b)$ says that in time $t$, information (as measured by the commutator of the initially localized observables) propagates at most a distance of order $t^a$, up to errors decaying like $x^{-b}$ away from the modified ``light cone'' $t^a=k$ in spacetime. The case $a=1$ corresponds to ballistic transport.

We will prove that the system exhibits the following features suggesting a transition from transport to localization as $\lam$ is increased. The precise statements are given later.
\be{itemize}
\item When $\lam$ is \emph{small enough}, PLR$(a,b)$ cannot hold if $a$ is too big or $b$ is too small. In other words, there exist observables $A,B$ for which the bound \eqref{eq:polyLR} fails and in this sense transport is at least of order $t^a$. To give a concrete example, we show that for $\lam<2$, \eqref{eq:polyLR} fails with probability one if $0\leq a\leq 1$ and $b>1/2$ satisfy
$$
a\l(1+\frac{1}{2b-1}\r)<1.
$$
In particular, for any $b>1/2$ there exists $a>0$ such that \eqref{eq:polyLR} does not hold. 

\item When $\lam$ is \emph{large enough}, the system is ``polynomially localized'' in the sense that
\beq
\label{analogue}
\mathbb E\big[\sup_{t\in\R}\|[\tau_t^n(A),B]\|\big]\leq C \|A\| \|B\|  \l(\frac 1 k\r)^{\kappa \lambda^2 - 5/4}
\eeq
for a coefficient $0<\kappa\leq \frac{5}{16}$. This is a disorder-averaged version of PLR$(0,\kappa\lam^2-5/4)$ and may be understood as a \emph{zero-velocity PLR bound}. It is of course only effective when $\lam$ is large enough to guarantee $\kappa\lam^2-5/4>0$. 

One can also remove the expectation and instead obtain a statement that holds with probability one by a simple interchange of summation, but this comes at the price of having the constant $C$ depend on the realization of the randomness.
\e{itemize}

\be{rmk}
\label{caveat}
\be{enumerate}[label=(\roman*)]
\item Of course, all statements are independent of the system size $n$ to retain relevance in the thermodynamic limit.
\item
It follows from \cite[Thm.\ 2.6]{DLLY} and Proposition \ref{betabound} that if only exponentially small errors are tolerated in an LR bound, then our model will exhibit ballistic transport for all $\lam>0$. This fits with the localization being only polynomial in type, even for large $\lam$.
\item 
We emphasize that our results do \emph{not} exclude that for small $\lam$, an analogue of \eqref{analogue} holds with the exponent $\kappa \lambda^2 - 5/4$ replaced by a number $b\leq 1/2$. If this were true, it would be misleading to speak of a true transition from non-trivial transport to localization and it is for this reason that we do not claim to prove such a transition.
\e{enumerate}
\e{rmk}

To prove the results, we use the standard method of expressing the $XY$ chain in terms of free fermions via the Jordan-Wigner transformation. The basic idea is to take bounds for the corresponding one-body system \cite{MR797277, MR2097423, MR1741780, MR1628290} and to pull them through the (non-local) Jordan-Wigner transformation by extending ideas of \cite{MR2966945}. 

\cite{MR2966945} considered a non-decaying random external field which yields an exponentially localized system. We extend their ideas to a situation in which errors decay only polynomially. Related papers which study the dependence of paramters in LR bounds and their generalizations on the external field are \cite{DLLY2,DLLY, DLY, K}. The idea of studying polynomial LR bounds was conceived in \cite{DLLY2,DLLY}, but there it was only shown that the idea does \emph{not} apply to the random dimer model (a model with anomalous one-body transport).

For large $\lam$, we use the fact that the Kunz-Souillard method utilized in \cite{MR797277} actually yields a polynomial bound on the eigenfunction correlator \eqref{j}. We are grateful to David Damanik for pointing us to \cite{MR797277}.\\

As mentioned before, we also show similar results for particle number transport. We have also attempted without success to prove analogous results for the entanglement entropy of eigenstates in the spirit of the recent works \cite{ANSS,AS,EPS,PS}. However we ran into difficulty bounding the entanglement entropy of eigenstates in the ``localization regime'' of large $\lam$ because of the growth in $j$ of the bound \eqref{j}. We believe that this question constitutes an interesting open problem.

We close the introduction with

\be{rmk}
For the PLR$(a,b)$ bounds defined by \eqref{eq:polyLR} and \eqref{analogue}, we only consider observables $A$ supported at site $1$. If $A$ is supported at a site $j>1$, the decaying factor is \emph{not} replaced by the distance of the supports $|j-k|$ (as would be the case in a direct polynomial generalization of the LR bound, compare \cite{DLLY2,DLLY}), but instead by $\min\{j,k\}/\max\{j,k\}$. The precise statement is in Theorem \ref{thm:exp-LR}. The reason why one cannot expect the distance $|j-k|$ is that the system is far from being translation-invariant. 
\e{rmk}

\section{The model}

\subsection{The XY Chain in a random decaying external field}
For every $n\in\N=\{1,2,3,\ldots\}$, we consider the Hilbert space 
\beqs
\mathcal{H}_n= \bigotimes_{j=1}^n \set C^2.
 \eeqs
On $\mathcal H_n$, the Hamiltonian of the isotropic XY chain with a random decaying external field is given by 
\beqs
    H_{n}^{XY}(\omega):= - \sum_{j=1}^{n-1} \l(\sigma^x_j \sigma^x_{j+1} +\sigma^y_j \sigma^y_{j+1}   \r) +\lambda\sum_{j=1}^n \frac {V_j(\omega)}{j^{1/2}}  \sigma^z_j,
\eeqs
where $\lambda>0$ is a coupling constant. 
The sequence $\big(V_{j}(\omega)\big)_{j\in\N}$ is a family of i.i.d. random variables on a probability space $(\Omega, \Sigma, \mathbb P)$. We assume that its single-site distribution has zero mean and is absolutely continuous with a bounded density of compact support. 
In the above,
\beqs
    \sigma^x =\ttmatrix{0}{1}{1}{0},\quad \sigma^y =\ttmatrix{0}{-i}{i}{0},\quad \sigma^z =\ttmatrix{1}{0}{0}{-1}
\eeqs
are the \emph{Pauli matrices}
and $\sigma^{x,y,z}_j$ is short-handed for
\beqs
\ind_1\otimes\ldots \ind_{j-1} \otimes \sigma^{x,y,z}\otimes \ind_{j+1}\ldots \otimes \ind_n
\eeqs
 for $1\le j \le n$. 
 In the following we omit the $\omega$-dependence for brevity.  
For a finite set $J\subset \N$, we define the algebra of observables supported on $J$ by
\beqs
 \curly{A}_J = \bigotimes_{j\in J} \curly{B}(\set C^2),
\eeqs
where $\curly{B}(\set C^2)$ is the set of all complex $2\times 2$ matrices. We will often make use of the fact that for $J\subset J'$, there is a natural embedding of $\curly{A}_J$ into $\curly{A}_{J'}$ by tensoring with the identity on $J'\setminus J$. Also, we set $\curly{A}_j\equiv \curly{A}_{\{j\}}$.

Finally, the \emph{Heisenberg dynamics} of an observable $A\in \curly{A}_J$ under the Hamiltonian $H_n^{XY}$ is defined by
\beq
\label{HBdefn}
    \tau_t^n(A) := e^{it H_n^{XY}} A e^{-it H_n^{XY}}.
\eeq

\subsection{The Jordan-Wigner transformation}

We use the standard procedure, going back to \cite{LIEB1961407}, of mapping the XY chain to free fermions via the Jordan-Wigner transformation.

For the details of the diagonalization procedure, we refer to Section~3.1 in \cite{MR2966945}. Here we only recall what we need to establish notation. The first step is to introduce the lowering operator
\beq\label{def:a}
 a_j = \frac{1}{2} \l(  \sigma_j^x-i \sigma^y_j  \r) = \ttmatrix{0}{0}{1}{0}_j
\eeq
and its adjoint the raising operator $a_j^*$ for all $1\leq j\leq n$. The Jordan-Wigner transformation maps these to the fermion operators
\beq
\label{eq:JW}
    c_1 = a_1,\quad c_j = \sigma^z_1\ldots \sigma^z_{j-1} a_j\quad \text{for } 2\leq j\leq n.
\eeq
The $\{c_j\}$ then satisfy the canonical anticommutation relations (CAR). We have the identity 
\beq\label{a*a=c*c}
a^*_ja_j = c^*_j c_j.
\eeq
In terms of the fermion operators, the Hamiltonian reads, 
\beq
\label{Hrewrite}
    H^{XY}_n = 2 \mathcal C^* H_n \mathcal C -\sum_{j=1}^n \tilde V_j
\eeq
where $\mathcal C:= ( c_1,...,c_n)^T$ and $\widetilde V_{j}:=\frac \lambda {j^{1/2}} V_j$. The $n\times n$ matrix $H_n$ is given by 
\beq
\label{eq:Hndefn}
    H_{n}=\left(\be{array}{cccc}
    \widetilde V_1\ & 1 & &\\
    1 & \ddots & \ddots & \\
     & \ddots & \ddots & 1 \\
    & &1 & \widetilde V_n\\
    \e{array}	\right),
\eeq
Note that $H_{n}$ can be identified with a discrete Schr\"odinger operator on the half line, i.e. on $\ell^2(\N)$, with the random decaying potential $\{\tilde V_j\}$ and zero boundary conditions at site $n+1$. The constant $\sum_{j=1}^n \tilde V_j$ in \eqref{Hrewrite} does not change the Heisenberg dynamics \eqref{HBdefn} and can thus be ignored in the following.

We will often use that the Heisenberg dynamics of the $c_j$ operators is given in the following simple fashion.

\be{prop}\cite[Sec. 3]{MR2966945}
\label{prop:tau}
For all $1\leq j,k\leq n$, the identity
\beq
\label{eq:3.15}
 \tau_t^n(c_j) = \sum_{m=1}^n \<\delta_j,e^{-2itH_{n}}\delta_m\> c_m
\eeq
holds and consequently
\beq
\label{eq:3.16}
 \|[\tau_t^n(a_j), B]\| \le 
  2\sum_{m=1}^n| \<\delta_j,e^{-2itH_{n}}\delta_m\>| \l(\|[ c_m, B]\|+\|[ c_m^*, B]\|\r)
\eeq
\e{prop}

\be{proof}
\eqref{eq:3.15} follows from diagonalizing the one-particle operator $H_n$, see \cite[Eq. (3.15)]{MR2966945}.
Taking adjoints, the same is also true for $c_k^*$.  The bound \eqref{eq:3.16} follows directly from \cite[Eq. (8)]{DLLY} by applying \eqref{eq:3.15}.
\e{proof}

\section{Polynomial Lieb-Robinson type bounds}

\subsection{Localization for large enough $\lam$}

We start with recalling an old result by \cite{MR797277} which provides bounds on the eigenfunction correlator of the Anderson model with a random decaying potential.

\be{lm}\label{lemma1}
Let $H_n$ be the operator given in \eqref{eq:Hndefn}. Then there exist constants $C,\kappa>0$ such that for all $n\in\N$ and all $1\leq j\le k\leq n$, we have
\beq
\label{j}
\mathbb E\big[\sup_{|g|\leq 1} |\<\delta_j,g(H_n)\delta_k\>| \big] \leq \frac{C}{\lambda}(jk)^{1/4}\left(\frac j { k}\right)^{ \kappa\lambda^2}. 
\eeq
\end{lm}

In particular, one can choose $g(x)= e^{-itx}$ in the above. The exponent $\kappa$ will feature in all of the following bounds and we show later that it satisfies $\kappa\leq \frac{5}{16}$, see Corollary \ref{kappabound}.

\begin{proof}
We estimate
\beq
\mathbb E \big[ \sup_{|g|\leq 1} |\<\delta_j,g(H_n)\delta_k\>| \big]
\leq 
\E\Big[\sum_{E\in\sigma(H_n)} |\psi_E^n(j)||\psi_E^n(k)|\Big]=:\overline \rho^n(j,k,\R )
\eeq
where the sequence $\big(\psi^n_E\big)_{E\in \sigma(H_n)}$ denotes the normalized eigenvectors of $H_n$ counted with multipicity.
An adaption of \cite[Prop. III.1]{MR797277} implies
\beq
\overline\rho^n(j,k,\R )\le \frac{C}{\lambda}(jk)^{1/4}\left(\frac j { k}\right)^{ \kappa\lambda^2}.
\eeq
The latter follows from inequality \cite[Eq. III.16]{MR797277} using the bounds \cite[Eq. III.14 and eq. III.15]{MR797277} and we remark that in the result \cite[Eq. III.4]{MR797277} the $1/2$-exponent should be replaced by an $1/4$-exponent.
\end{proof}

As a consequence, we obtain a disorder-averaged polynomial Lieb-Robinson type bound with $a=0$ for the spin chain $H_n^{XY}$. 

\be{thm}\label{thm:exp-LR}
Let $\kappa$ be as in Lemma \ref{lemma1} above. Suppose that $\kappa\lam^2>\frac{5}{4}$. Then there exists a constant $C>0$ such that for all choices of $1\leq j\leq k\leq n$, 
\beq\label{L-R-bound}
\mathbb E\big[\sup_{t\in\R}\|[\tau_t^n(A),B]\|\big]\leq C\|A\| \|B\|   (j k)^{5/4} \l(\frac j k\r)^{\kappa\lambda^2 }
\eeq 
holds for all observables $A\in \mathcal A_j$ and $B\in \mathcal A_{k,...,n}$. 
\end{thm}

We emphasize that the constant $C$ is also uniform in $n$.

\be{proof}
Note that $\A_j$ is spanned by the matrices $\{ a_j,a^*_j, a_ja^*_j, a^*_j a_j\}$.
According to Proposition \ref{prop:tau}, we can estimate 
\beq
\label{eq1}
\mathbb E \| [ \tau_t^n(a_j),B] \|
\le 
  2\sum_{m=1}^n| \<\delta_j,e^{-2itH_{n}}\delta_m\> |\l(\|[ c_m, B]\|+|\|[ c_m^*, B]\|\r)
\eeq
We note that $[c_m,B]=0$ for all $m<k$.
Hence, Lemma \ref{lemma1} implies
\beq
\be{aligned}
\eqref{eq1}
\le &
\frac{4C}{\lambda}\| B\|   \sum_{l=1}^j \sum_{m=k}^n 
 (lm)^{1/4}\left(\frac l { m}\right)^{ \kappa\lambda^2}
\\
\le & \frac{4C}{\lambda}\| B\|   \sum_{l=1}^j \sum_{m=k}^\infty 
 (lm)^{1/4}\left(\frac l { m}\right)^{ \kappa\lambda^2}
  \\
\le &
\frac{C}{\lambda}\| B\| (j k)^{5/4}  \l(\frac j k\r)^{\kappa\lambda^2}
\end{aligned}
\eeq
for some constant $C>0$ which is finite for $\lam>\sqrt{\frac{5}{4\kappa}}$. Taking adjoints the same estimate is true for $a^*_j$.
For the products $a_j^*a_j$ and $a_ja^*_j$, we use the Leibniz rule $[AB,C]=A[B,C]+[A,C]B$.
\end{proof}

\be{rmk}
Instead of the distance $|j-k|$ of the supports of the observables, which would appear in a straightforward polynomial generalization of the traditional LR bound as was proposed in \cite{DLLY2,DLLY}, the right hand side depends on the quotient $j/k$. Note that the distance $|j-k|$ is not so natural for our model, which is far from being translation-invariant.
 
However, if we consider observables $A$ supported at a \emph{fixed} site, say the site $1$, the bound \eqref{L-R-bound} reduces to a polynomial Lieb-Robinson bound involving the distance of the supports. Let $A\in \A_1$. Then the bound
\beq
\mathbb E\big[\sup_{t\in\R}\|[\tau_t^n(A),B]\|\big]\leq C \|A\| \|B\|  \l(\frac 1 k\r)^{\kappa\lambda^2 - 5/4}
\eeq
holds uniformly in $n\in\N$ and $B\in \mathcal A_{k,...,n}$ for some $1\leq k\leq n$.
\end{rmk}

For small $t$ the above is not satisfactory. One can improve the result:

\be{prop}
Let $\kappa$ be as in Lemma \ref{lemma1}. There exists a constant $C$ such that for all choices of $1\leq j\leq k\leq n$,
\beq
\mathbb E\l[ \|[\tau_t^n(A),B]\|\r]\leq C \|A\| \|B\|  |t| \Big(\frac 1 {k} \Big)^{\kappa \lambda^2 -5/4 }
\eeq 
holds for all observables $A\in\A_1$,  $B\in \mathcal A_{k,...,n}$.
\end{prop}

\be{proof}
We follow the proof of \cite[Cor. 3.4]{MR2966945}. Define
\beq
f(t):=[\tau_t(A),B].
\eeq
Then, $f(t)$ solves the ODE
\beq
f'(t)=i[f(t), \tau_t^n(H_1)]- i[[B,\tau_t^n(H_1)],\tau_t^n(A)].
\eeq
where $H_1:= \sigma_1^x\sigma_2^x+ \sigma_1^y\sigma_2^y + V_1\sigma_1^z$. Following \cite[App. A]{MR2256615} we obtain
\beq
\| f(t)\| \le \int_0^{|t|} \d s\, \| [\tau_s^n(H_1),B] \|.
\eeq
Since $H_1$ is supported on $\mathcal A_1\otimes\mathcal A_2$ we use Theorem \ref{thm:exp-LR} to obtain a time independent bound on the integrand which yields the theorem. 
\end{proof}

Instead of bounds in expectation, one can also obtain almost sure bounds, but at the price of getting an $\om$-dependent constant. 

\be{cor}
Let $\lambda$ be big enough. Let $A\in \A_1$ and $B\in \mathcal A_{k,...,n}$.
For all $\epsilon>0$ and $\mathbb P$-almost every $\omega$, there exists a random constant $C_\omega^\eps(A,B)$ independent of $n$ such that 
\beq
\sup_{t\in\R}\|[\tau_t^n(A),B]\|\leq C^\epsilon_\omega(A,B)
 \left( \frac 1 {k}\right)^{\kappa\lambda^2-5/4-\epsilon}.
\eeq 
\end{cor}

\be{proof}
For any $\epsilon>0$
\beq
\sum_{k\in\mathbb N} 
\mathbb E\big[\sup_{t\in\R}\|[\tau_t^n(A),B]\|\big]
|k|^{\kappa\lambda^2-9/4-\epsilon}<\infty.
\eeq
The claim now follows from Fubini's theorem.
\e{proof}

Note that the exceptional set of $\om$'s depends on $A$ and $B$.

\subsection{Lower bounds on transport for small enough $\lam$}

In this section we restrict ourselves to pairs of observables for which one of the observables is supported at the site $1$. 
\be{defn}
Let $0\leq a\leq 1$ and $b\geq 0$. We say that $H^{XY}_{n}$ exhibits the polynomial Lieb-Robinson type bound PLR$(a,b)$, if there exists a constant $C>0$ such that for all $n\in\N$
\beq\label{low:eq1}
\| [ \tau_t^n(A),B]\| \le C \l( \frac {t^a}{k}\r)^b 
\eeq
holds for all $A\in\mathcal A_1$, $B\in\mathcal A_{k,...,n}$. 
\end{defn}

Let $H$ be the discrete Schr\"odinger operator on $\ell^2(\N)$ which arises as the inductive limit of the family $(H_n)_{n\in\N}$. 

\be{defn}
We define the $p$-th moment of the position operator
\beq
\label{Xdefn}
    |X|^p(t):=\sum_{k\in\N} k^p |\scp{e^{-itH}\de_j}{\de_k}|^2
\eeq
and its \emph{time-average}
\beq
 \langle |X|^p\rangle(T) := \frac{2}{T}   \int_0^\it \d t\, e^{-2t/T}|X|^p(t)
\eeq
for all $T>0$. The upper and lower transport exponents are defined by
\beq
\beta^-(p):=\liminf_{t\to\infty} \frac{\ln |X|^p(t)}{p\ln t}
\quad\text{and}\quad
\beta^+(p):=\limsup_{t\to\infty} \frac{\ln |X|^p(t)}{p\ln t}
\eeq
and their averaged versions are defined by
\beq
\langle \beta^-(p)\rangle := \liminf_{T\to\infty} \frac{\ln \< |X|^p\>(T)}{p\ln T}
\quad\text{and}\quad
\langle \beta^+(p)\rangle := \limsup_{T\to\infty} \frac{\ln \< |X|^p\>(T)}{p\ln T}.
\eeq
\end{defn}

\be{thm}\label{thm1}
Assume \textnormal{PLR}$(a,b)$ holds for some $0\leq a\leq 1$ and $b>1/2$. 
Then,
\beq\label{low:eq2}
\limsup_{\epsilon\to 0}\beta^+(2b-1-\epsilon) \le a\l(1+ \frac 1 {2b-1} \r).
\eeq
\end{thm}

\be{proof}
The strong resolvent-convergence of $H_n$ to $H$ (this follows e.g. from the geometric resolvent identity) 
implies the convergence
\beq
\lim_{n\to\infty} \< e^{itH_n}\delta_1,\delta_k\>  =  \< e^{itH}\delta_1,\delta_k\>,
\eeq
for any $1\le k \le n$.
Hence,  Fatou's lemma implies the inequality
\be{align}
\sum_{k\in \N} k^{2b-1-\epsilon} |\<e^{-itH}\delta_1,\delta_k\>|^2
= & 
\lim_{M\to\infty}
\sum_{1\le k\le M} k^{2b-1-\epsilon} |\<e^{-itH}\delta_1,\delta_k\>|^2\nonumber\\
 \le &
\lim_{M\to\infty}\liminf_{n\to\infty}
\sum_{1\le k\le M} k^{2b-1-\epsilon} |\<e^{-itH_n}\delta_1,\delta_k\>|^2,\label{beta:eq1}
\end{align}
where $\epsilon>0$ is arbitary.

Now, we bound the one-body propagation in terms of the many-body propagation using \cite[Lm. 4.1]{DLLY}. It implies that for any $1\le k\le n$
\beq
|\<e^{-itH_n}\delta_1,\delta_k\>|\le \| [ \tau_t^n(c_1), a_k^* ] \|.
\eeq
Using this and the assumption that PLR$(a,b)$ holds, we bound
\be{align}
\eqref{beta:eq1}
 \le & \
t^{2ab} \sum_{k\in \N} k^{-1-\epsilon}.
\end{align}
Since the latter is summable for any $\epsilon>0$, this implies
\beq
\beta^+(2b-1-\epsilon) \le \frac{2ab}{2b-1-\epsilon}
\eeq
and therefore \eqref{low:eq2} follows.
\end{proof}


\begin{prop}\label{betabound}
Let $p>\frac \lambda 4$. The lower bound
\beq\label{betaeq1}
\beta^+(p) \ge 1- \frac{\lambda}{4p}
\eeq
holds $\mathbb P$-almost surely. In the case of $\lambda<2$ one has 
\beq \label{betaeq2}
\beta^+(p) = 1
\eeq
$\mathbb P$-almost surely.
\end{prop}

Before we give the proof, which is based on results in \cite{MR2097423,MR1741780,MR1628290}, we discuss the consequences of combining Theorem \ref{thm1} and Proposition \ref{betabound}. What we obtain can be interpreted as lower bounds on transport, as we explained in the introduction, however see also the caveat in Remark \ref{caveat}(iii).

\be{cor}
Let $(a,b)$ be a pair of $0\leq a\leq 1$ and $b>1/2$. If either of the following two conditions applies, then \textnormal{PLR}$(a,b)$ cannot hold.
\be{itemize}
\item $\lambda<2$ and $a\l(1+ \frac 1{2b-1}\r) < 1$
\item $2\leq \lambda< 2b-1$ and $a\l(1+ \frac 1{2b-1}\r) < 1- \frac \lambda {4(2b-1)}$.
\e{itemize}
In particular, for any fixed pair $(a,b)$ of $0\leq a\leq 1$ and $b>1/2$, one can choose $\lam$ large enough to get that PLR$(a,b)$ cannot hold.
\e{cor}

\be{rmk}
A shortcoming of our results is that we need to assume $b>1/2$, see Remark \ref{caveat}(iii). This is ultimately a consequence of summing up one-body transport bounds when inverting the Jordan-Wigner transformation (compare Proposition \ref{prop:tau}) and is therefore intimately connected to the core of the method.
\e{rmk}

We also get a bound on the maximal power of the polynomial decay coefficient $\kappa$ which was introduced considered in the previous section.

\be{cor}\label{kappabound}
The constant $\kappa$ from Proposition \ref{lemma1} satisfies $\kappa\leq \frac{5}{16}$.
\e{cor}

\be{proof}
Note that $\kappa$ is independent of $\lam$. Fix $\lam<2$ and $p>0$. By Proposition \ref{betabound}, $\sup_{t>0}|X|^p(t)=\it$. Recalling the definition \eqref{Xdefn} of $|X|^p(t)$ and using the estimate in Lemma \ref{lemma1} then gives $p+1/4-\kappa\lam^2\geq -1$. Sending $\lam \to 2$ and $p\to 0$ yields $\kappa\leq \frac{5}{16}$.
\e{proof}

It remains to give the

\be{proof}[Proof of Prop. \ref{betabound}]
For equation \eqref{betaeq1}, we apply the lower bound \cite[Thm. 5.1, Eq. (5.3)]{MR2097423} to the function $f\in C_c^\infty(\R)$ with $f\equiv 1$ on $\sigma(H)$. This provides for any $\eps>0$ the bound
\beq
\<|X|\>_j^p(T) \ge C_\omega(p,\eps) T^{p-2\gamma-\eps},
\eeq
$\mathbb P$-almost surely, 
where $\gamma := \inf_{E\in (-2,2)} \frac \lam {8-2 E^2}$. This implies
\beq\label{eq1111}
\langle \beta^-(p)\rangle \ge 1 - \frac{\lam}{4p}.
\eeq
The chain of inequalities $\langle \beta^-(p)\rangle  \le \langle \beta^+(p)\rangle\le \beta^+(p)$ gives the result. To see the last inequality, note that $\beta:=\beta^+(p)>0$ implies for any $ \epsilon>0$, $|X|^p_1(t) \le C t^{p\beta  + \epsilon}$. This readily gives
\beq
\<|X|_1^p\>(T)= \frac 2 T \int_0^\infty \d t\, e^{-2t/T} |X|_1^p(t) \le C T^{p\beta +\epsilon}
\eeq
and the inequality $\langle \beta^+(p)\rangle\le \beta$.

For equation \eqref{betaeq2}, we use \cite[Thm 5.1]{MR1741780} with $m=p$, where we have to prove its assumption, which is $P_c\delta_1\neq 0$. Here, $P_c$ is the orthogonal projection onto continuous part of the spectrum.  Since $|\lambda|<2$, the operator $H$ exhibits singular continuous spectrum \cite{MR1628290}, thus $P_c\neq 0$. Now, $P_c\delta_1\neq 0$ follows from cyclicity of $\delta_1$, which can be proven by induction because the Hamiltonian acts on the half space $\ell^2(\N)$ only. 
\end{proof}

\section{Propagation bounds for the number operator}
In this section, we derive bounds on the propagation of the number operator. These follow easily by combining a computation in \cite{ANSS} with the bounds on the one-body dynamics discussed before.

We define the number operator and the local number operator by
\beq
\mathcal N := \sum_{j=1}^n a_j^*a_j\quad\text{and}\quad \mathcal N_S := \sum_{j\in S} a_j^*a_j ,
\eeq
where $a_j$ is given in \eqref{def:a} and $S\subset \{1,...,n\}$.
This measures the number of up-spins in $S$. 
Let
\beq
\label{rhodefn}
\rho= \bigotimes_{j=1}^n \rho_j,\qquad \rho_j:= 
\be{pmatrix}
\eta_j & 0 \\ 0 & 1-\eta_j
\e{pmatrix}
\eeq
and $0\leq \eta_j \le 1$. We denote by $\rho_t:= e^{-itH_n} \rho e^{itH_n}$ the time evolution of the state $\rho$ and by
$\< A\>_{\rho}:=\tr { A \rho }$ the expectation of an observable $A$ with respect to the state $\rho$. 
\begin{thm}
Let $\kappa>0$ be as in Lemma \ref{lemma1}. There exists a constant $C>0$ such that for every $n\geq 1$ and $S\subset \{1,\ldots,n\}$, 
\beq
\E \Big[ \sup_{t\geq 0} \< \mathcal N_S \>_{\rho_t}\Big] \leq 
\frac{C}{\lambda}
\sum_{j\in S} \sum_{k=1}^n \eta_k
(jk)^{1/4}\left(\frac {\min\{j,k\}} {\max\{j, k\}}\right)^{ \kappa\lambda^2} .
\eeq
\e{thm}

This follows directly by combining results of \cite{ANSS} with Lemma \ref{lemma1}.

\be{rmk}
To illustrate the above we split $\{1,...,n\}= I\cup J$ with $I:=\{1,...,m\}$ and $J:=\{m+1,...,n\}$ for  $n>m\in \N$. We set $\eta_j=0$ on $I$ and $\eta_j=1$ on the complement $J$. In other words $\rho= |\varphi\>\<\varphi|$ with the vector
\beq
|\varphi\> = |\downarrow\>^{\otimes m} \otimes |\uparrow\>^{\otimes(n-m+1)}
\eeq
in standard notation. Let $m>l\in \N$ and $S=\{1,...,l\}$. For $\kappa\lambda^2> 5/4$, the above theorem implies the bound
\beq
\E \Big[ \sup_{t\geq 0} \< \mathcal N_S \>_{\rho_t}\Big] \leq  C\l(\frac l m \r)^{\kappa\lambda^2} (lm)^{5/4} 
\eeq
for a constant $C>0$ uniform in $l,m,n$. 
This is a time-independent bound on the number of up-spins which propagate from $J$ into $S$ and it decays as the distance $m\to\infty$ (when $\lam$ is large enough to guarantee $\kappa\lambda^2> 5/4$).
\e{rmk}

\be{proof}
The same computation that gives \cite[eq.\ (41)]{ANSS} shows
\beq
\label{eq:41}
\<\mathcal N_S\>_{\rho_t}
= 
\sum_{j\in S} \sum_{k=1}^n |\<\delta_j, e^{2itH_n} \delta_k\>|^2 \eta_k.
\eeq
Using this, Lemma \ref{lemma1} implies
\beq
\be{aligned}
\E\Big[ \sup_{t\ge 0} \<\mathcal N_S\>_{\rho_t} \Big]
\leq 
\sum_{j\in S} \sum_{k=1}^n  \eta_k \E\Big[\sup_{t\geq 0} |\<\delta_j, e^{2itH_n} \delta_k\>|^2 \Big]
\e{aligned}
\eeq
The assertion now follow from $|\<\delta_j, e^{2itH_n} \delta_k\>|^2\leq |\<\delta_j, e^{2itH_n} \delta_k\>|$ and  Lemma \ref{lemma1}. 
\e{proof}
 
 
\be{thm}
\label{thm:PNLB}
If for some $0\leq a\leq 1<b$ and all $k,n\in\N$ with $k\leq n$
\beq\label{numbtransassump}
 \<\mathcal N_1\>_{\rho_t} \leq \l(\frac {t^a}k\r)^b
\eeq
holds for all $\rho$ of the form \eqref{rhodefn} and $\eta_j=0$ for $j<k$.
Then, the upper transport exponent satisfies the bound
\beq
\limsup_{\eps\to 0}\beta^+(b - 1-\eps) \le \frac{ab} {b-1}.
\eeq
\e{thm}

Again, Proposition \ref{betabound} then gives restrictions on the possible values of $0\leq a\leq 1<b$ for which \eqref{numbtransassump} can hold. Therefore Theorem \ref{thm:PNLB} may be interpreted as a lower bound on the transport of particles (from sites $k$ and larger to the site $1$) if at most error of order $x^{-b}$ with $b>1$ can ignored, compare Remark \ref{caveat}(iii).

\be{proof}
Let $\rho_k$ be given as in \eqref{rhodefn} with $\eta_j=\delta_{j,k}$. By  \eqref{eq:41}
\beq
\< \mathcal N_1\>_{\rho_t^k}=|\<\delta_1, e^{-itH_n} \delta_k\> |^2.
\eeq
Hence, the computation in \eqref{beta:eq1} and assumption \eqref{numbtransassump} imply that for any $p>0$
\beq
\be{aligned}
|X|^p(t) \leq &\lim_{M\to\infty} \liminf_{n\to\infty} \sum_{1\le k\le M} k^p|\< e^{-itH_n}\delta_1,\delta_k\>|^2\\
\leq &
 \sum_{k\in\N} k^p \l(\frac {t^a}k\r)^{b}
 = t^{ab} \sum_{k\in\N} k^{p-b}.
\e{aligned}
\eeq
Taking $p= b - 1-\eps$ for an $\eps>0$, the last sum is finite and this gives the assertion. 
\e{proof}

\section*{Acknowledgements}
We thank David Damanik for pointing us to reference \cite{MR797277}. M.G. is grateful to Gian Michele Graf for his kind hospitality at {ETH} {Z\"urich}.


\newcommand{\etalchar}[1]{$^{#1}$}

\end{document}